In this paper we prove the following: (1) The basic error of time-dependent perturbation theory is using the sum of first finite order of perturbed solutions to substitute the exact solution in the divergent interval of the series for calculating the transition probability. In addition quantum mechanics neglects the influence of the normality condition in the continuous case. In both cases Fermi's golden rule is not a mathematically reasonable deductive inference from the Schrödinger equation. (2) The transition probability per unit time deduced from the exact solution of the Schrödinger equation is zero, which cannot be used to describe the transition processes.

\\

## 1. THE MOST IMPORTANT UNSETTLED QRESTION IN PHYSICAL THEORY

Quantum mechanics deems that we can use the Schrödinger Equation

$$i\hbar \frac{da_f(t)}{dt} = \sum_k {}^{(S)}H_{fk}'g(t)\exp\left[\frac{i(E_f - E_k)t}{\hbar}\right]a_k(t), \tag{1}$$

to describe that the state varies with time. According to the time-dependent perturbation theory, suppose that

$$a_f(t) = \sum_{N=0}^{\infty} a_f^{(N)}(t) \tag{2}$$

and use the approximate equations

$$i\hbar \frac{da_f^{(0)}(t)}{dt} = 0, \tag{3}$$

$$i\hbar \frac{da_f^{(N)}(t)}{dt} = \sum_k {}^{(S)}H_{fk}' g(t)\exp\left[\frac{i(E_f - E_k)t}{\hbar}\right]a_k^{(N-1)}(t), \qquad if \qquad N \geq 1 \tag{4}$$

to substitute (1). This set of equations is so complex, people only find the first finite order approximate solutions, and none finds the approximate solution $a_f^{(N)}(t)$ for arbitrary $N$ order. Then people can only take the sum of first $M$ orders approach

$$\sum_{N=0}^{M} a_f^{(N)}(t)$$

to substitute the exact solution $a_f(t)$. If the difference between the exact solution and the sum of first $M$ orders approach is large than the experimental error, then the comparison between this sum and the experimental value is meaningless. In this case we cannot obtain the conclusion that the Schrödinger equation may used to describe the transition processes. Only in the convergent range we have that, for each $e > 0$, there is a number $M$ that the relation

$$\left| a_f(t) - \sum_{N=0}^{M} a_f^{(N)}(t) \right| = \left| \sum_{N=M+1}^{\infty} a_f^{(N)}(t) \right| < e \qquad (5)$$

is satisfied. What is $a_f^{(N)}(t)$ for arbitrary $N$ order approach? What is the exact solution? In which range is the series convergent? This set of question is relative to that problem, can we use the Schrödinger equation to describe the transition processes. So it is the most important unsettled question in physical theory.

According to quantum mechanics, we shall discuss the questions in the cases, that is the time factor is

$$g_1(t) = \begin{cases} 1, & if \qquad 0 \le t \le \infty \\ 0, & if \qquad t < 0 \end{cases} \qquad (6)$$

or

$$g_2(t) = \begin{cases} \cos(\omega\, t + \delta) & if \qquad 0 \le t \le \infty \\ 0 & if \qquad t < 0 \end{cases} \qquad (7)$$

and the eigenvalues of $\hat{H}_0$ are discrete or continuous.

# 1. DISCUSSION OF THE CASE IN WHICH THE TIME FACTOR IS $g_1(t)$ AND EIGENVALUES OF $\hat{H}_0$ ARE DISCRETE

Now we must (i) find out $a_f^{(N)}(t)$ for arbitrary N; (ii) deduce $a_f(t)$; (iii) calculate the transition probability per unit time from $a_f(t)$; (iv) compare the result with the transition probability per unit time deduced the first finite-order approximation.

## 2.1 Expression of the approximate solution $a_f^{(N)}(t)$ for arbitrary N

In the discrete case the initial condition is

$$a_f^{(N)}(0) = \begin{cases} \delta_{fi} & if \qquad N = 0 \\ 0 & if \qquad N > 0 \end{cases} \qquad (8)$$

As we know, the zeroth and first-order approximations of the time-dependent perturbed approximation are

$$a_f^{(0)}(t) = \delta_{fi,} \tag{9}$$

$$a_f^{(1)}(t) = \begin{cases} \dfrac{{}^{(S)}H'_{fi}}{E_i - E_f}\left\{\exp\left[\dfrac{i(E_f - E_i)t}{\hbar}\right] - 1\right\}, & \text{if } f \neq i \\[2mm] {}^{(S)}H'_{ii}\left[\dfrac{-it}{\hbar}\right], & \text{if } f = i \end{cases} \tag{10}$$

They can be rewritten as

$$a_f^{(0)}(t) = \sum_l D_{fi}^{(0)}(l,0)\exp\left[\frac{i(E_f - E_l)t}{\hbar}\right] \tag{11}$$

$$a_f^{(1)}(t) = \sum_l \left[D_{fi}^{(1)}(l,0) + D_{fi}^{(1)}(l,1)\left(\frac{-it}{\hbar}\right)\right]\exp\left[\frac{i(E_f - E_l)t}{\hbar}\right] \tag{12}$$

from which

$$D_{fi}^{(0)}(l,0) = \delta_{fl}\delta_{li} \tag{13}$$

$$D_{fi}^{(1)}(l,0) = c_f^{(1)}(l)\delta_{li} + c_i^{(1)}(l)^*\delta_{fl}$$

$$c_j^{(1)} = \begin{cases} {}^{(S)}H'_{jl}/(E_l - E_j) & l \neq j \\ 0 & l = j \end{cases} \tag{14}$$

$$D_{fi}^{(1)}(l,1) = {}^{(S)}H'_{ll}\,\delta_{fl}\delta_{li.} \tag{15}$$

Using the inductive method, we can prove that

$$a_f^{(N)}(t) = \sum_l \sum_{n=0}^{N} D_{fi}^{(N)}(l,n)\frac{1}{n!}\left[\frac{-it}{\hbar}\right]^n \exp\left[\frac{i(E_f - E_l)t}{\hbar}\right] \tag{16}$$

for arbitrary N, where $D_{fi}^{(N)}(l,n)$ is the time independent coefficient.

Equation (16) is correct for N=0,1. Suppose that it is correct for N. Therefore in the $g_1(t)$ case we have

$$a_f^{(N+1)}(t) = \left(\frac{-i}{\hbar}\right)\int_0^t dt_1 \sum_k {}^{(S)}H'_{fk}\exp\left[\frac{i(E_f - E_k)t_1}{\hbar}\right]a_k^{(N)}(t_1)$$

$$= \left(\frac{-i}{\hbar}\right)\int_0^t dt_1 \sum_l \sum_{n=0}^{N} \sum_k {}^{(S)}H'_{fk}\,D_{ki}^{(N)}(l,n)\frac{1}{n!}\left(\frac{-it_1}{\hbar}\right)^n \exp\left[\frac{i(E_f - E_l)t_1}{\hbar}\right]$$

$$\tag{17}$$

The result of this integral has a form analogous to (16). The question is that, what is the maximum of $n$? In (17), $l$ may takes all possible values; if $l = f$, the integrated function is a polynomial of the Nth power of $t$, so the maximum of $n$ in the expression of $a_f^{(N+1)}(t)$ is

$N+1$. Therefore, $a_f^{(N+1)}(t)$ may be expressed as the form

$$a_f^{(N+1)}(t) = \sum_l \sum_{n=0}^{N+1} D_{fi}^{(N+1)}(l,n) \frac{1}{n!}\left(\frac{-it}{\hbar}\right)^n \exp\left[\frac{i(E_f - E_l)t}{\hbar}\right] \tag{18}$$

Substituting (16) and (18) into (4), we get

$$\sum_l \sum_{n=0}^{N+1} D_{fi}^{(N+1)}(l,n)\left[(E_l - E_f)\frac{1}{n!}\left(\frac{-it}{\hbar}\right)^n + \frac{1}{(n-1)!}\left(\frac{-it}{\hbar}\right)^{n-1}\right]\exp\left[\frac{i(E_f - E_l)t}{\hbar}\right]$$

$$= \sum_l \sum_{n=0}^{N} \sum_k {}^{(S)}H'_{fk} D_{ki}^{(N)}(l,n)\frac{1}{n!}\left(\frac{-it}{\hbar}\right)^n \exp\left(\frac{i(E_f - E_l)t}{\hbar}\right) \tag{19}$$

Equation (19) is satisfied under these conditions:

$$(E_l - E_f)D_{fi}^{(N+1)}(l,N+1) = 0,$$

$$(E_l - E_f)D_{fi}^{(N+1)}(l,n) + D_{fi}^{(N+1)}(l,n+1) = \sum_k {}^{(S)}H'_{fk} D_{ki}^{(N)}(l,n). \quad for \quad n \le N \tag{20}$$

These are the recurrence formulas of the coefficient.

From the initial condition

$$a_f^{(N+1)}(0) = 0, \tag{21}$$

we get

$$\sum_l D_{fi}^{(N+1)}(l,0) = 0, \quad for \quad N \ge 0. \tag{22}$$

The conclusions are that, (i) if (16) is correct for $N$, then it is correct for $N+1$. However (16) is correct for $N = 0, 1$, then it is correct for arbitrary $N$; (2) the coefficient satisfies the recurrence formulas (20) and (22).

## 1.2 The Relation between the Coefficient in $a_f^{(N)}(t)$ and the Stationary-State Perturbed Approximation

Consider the equation of stationary state

$$\lambda c_f(\lambda) = \sum_k (E_k \delta_{fk} + {}^{(S)}H'_{fk})c_k(\lambda), \tag{23}$$

and the eigenfunction satisfies the orthonormality conditions

$$\sum_\lambda c_j(\lambda)c_i(\lambda)^* = \delta_{ji}, \tag{24}$$

$$\sum_j c_j(\lambda)c_j(\nu)^* = \delta_{\lambda\nu}. \tag{25}$$

The perturbations of $\pmb{\lambda}$, $c_f(\pmb{\lambda})$ are

$$\lambda = \sum_{N=0}^{\infty} \lambda^{(N)}, \tag{26}$$

$$c_f(\lambda) = \sum_{N=0}^{\infty} c_f^{(N)}(\lambda), \tag{27}$$

They satisfy the set of equation

$$(\lambda^{(0)} - E_j)c_j^{(0)}(\lambda) = 0 \tag{28}$$

$$(\lambda^{(0)} - E_j)c_j^{(N)}(\lambda) + \sum_{L=1}^{N} \lambda^{(L)} c_j^{(N-L)}(\lambda) = \sum_k {}^{(S)}H'_{jk} c_k^{(N-1)}(\lambda) \tag{29}$$

$$\sum_{\lambda} c_f^{(0)}(\lambda)c_i^{(0)}(\lambda)^* = \delta_{fi} \tag{30}$$

$$\sum_{\lambda} \sum_{L=0}^{N} c_f^{(L)}(\lambda)c_i^{(N-L)}(\lambda)^* = 0, \qquad if \qquad N > 0, \tag{31}$$

The set of (20) and (22) is analogous to the stationary-state perturbation equations.

Using the symbols from (28) to (31), we can rewrite (13) to (15) in the following form:

$$D_{fi}^{(0)}(l,0) = c_f^{(0)}(\lambda)c_i^{(0)}(\lambda)^* \Big|_{\lambda^{(0)}=E_l} \tag{32}$$

$$D_{fi}^{(1)}(l,0) = \left[ c_f^{(1)}(\lambda)c_i^{(0)}(\lambda)^* + c_f^{(0)}(\lambda)c_i^{(1)}(\lambda)^* \right] \Big|_{\lambda^{(0)}=E_l} \tag{33}$$

$$D_{fi}^{(1)}(l,1) = \left[ \lambda^{(1)} c_f^{(0)}(\lambda)c_i^{(0)}(\lambda)^* \right] \Big|_{\lambda^{(0)}=E_l} \tag{34}$$

Using the inductive method, we can prove that the coefficients can be expressed as

$$D_{fi}^{(N)}(l,n) = \sum_{l=n}^{N} \theta^{(L)}(\lambda,n)b_{fi}^{(N-L)}(\lambda) \Big|_{\lambda^{(0)}=E_l} \;, \quad for \quad 0 < n \le N \tag{35}$$

$$D_{fi}^{(N)}(l,0) = b_{fi}^{(N)}(\lambda) \Big|_{\lambda^{(0)}=E_l} \tag{36}$$

from which

$$b_{fi}^{(N)}(\lambda) = \sum_{k=0}^{M} c_f^{(K)}(\lambda)c_i^{(M-K)}(\lambda)^* \tag{37}$$

According to (29), we have

$$\sum_k {}^{(S)}H'_{fk} b_{ki}^{(N)}(\lambda) = (\lambda^{(0)} - E_f)b_{fi}^{(N+1)}(\lambda) + \sum_{M=1}^{N+1} \lambda^{(M)}b_{fi}^{(N+1-M)}(\lambda) \tag{38}$$

and

$$\sum_k {}^{(S)}H'_{fk} D^{(N)}_{ki}(l,n) =$$

$$= \left\{ \sum_{L=n}^{N} \theta^{(L)}(\lambda,n) \left[ (\lambda^{(0)} - E_f) b^{(N+1-L)}_{fi}(\lambda) + \sum_{M=1}^{N+1-L} \lambda^{(M)} b^{(N+1-L-M)}_{fi}(\lambda) \right] \right\} \Bigg|_{\lambda^{(0)}=E_l}$$

$$= (\lambda^{(0)} - E_f) \sum_{L=n}^{N+1} \theta^{(L)}(\lambda,n) b^{(N+1-L)}_{fi}(\lambda) \Big|_{\lambda^{(0)}=E_l} + \sum_{K=n+1}^{N+1} \left[ \sum_{L=n}^{K-1} \lambda^{(K-L)} \theta^{(L)}(\lambda,n) \right] b^{(N+1-K)}_{fi}(\lambda) \Big|_{\lambda^{(0)}=E_l}. \tag{39}$$

If $N = 0, 1,$ then (35), (36) are correct, and

$$\theta^{(1)}(\lambda,1) = \lambda^{(1)} \quad . \tag{40}$$

Suppose that (35), (36) are correct for N. From (20) and (22) we get

$$(E_l - E_f) D^{(N+1)}_{fi}(l, N+1) = 0, \tag{41}$$

$$(E_l - E_f) D^{(N+1)}_{fi}(l,n) + D^{(N+1)}_{fi}(l,n+1) =$$

$$= \left\{ (\lambda^{(0)} - E_f) \sum_{L=n}^{N+1} \theta^{(L)}(\lambda,n) b^{(N+1-L)}_{fi}(\lambda) + \sum_{K=n+1}^{N+1} \left[ \sum_{L=n}^{K-1} \lambda^{(K-L)} \theta^{(L)}(\lambda,n) \right] b^{(N+1-K)}_{fi}(\lambda) \right\} \Bigg|_{\lambda^{(0)}=E_l}. \tag{42}$$

$$(E_l - E_f) D^{(N+1)}_{fi}(l,0) + D^{(N+1)}_{fi}(l,1) =$$

$$= \left\{ (\lambda^{(0)} - E_f) b^{(N+1)}_{fi}(\lambda) + \sum_{M=1}^{N+1} \lambda^{(M)} b^{(N+1-M)}_{fi}(\lambda) \right\} \Bigg|_{\lambda^{(0)}=E_l}. \tag{43}$$

$$\sum_l D^{(N+1)}_{fi}(l,0) = 0. \tag{44}$$

The solutions of this set of equations are

$$D^{(N+1)}_{fi}(l,n) = \sum_{K=n}^{N+1} \theta^{(K)}(\lambda,n) b^{(N+1-K)}_{fi}(\lambda) \Big|_{\lambda^{(0)}=E_l} \tag{45}$$

$$D^{(N+1)}_{fi}(l,n+1) = \sum_{K=n+1}^{N+1} \theta^{(K)}(l,n+1) b^{(N+1-K)}_{fi}(\lambda) \Big|_{\lambda^{(0)}=E_l} \tag{46}$$

$$\theta^{(K)}(\lambda,n+1) = \sum_{L=n}^{K-1} \lambda^{(K-L)} \theta^{(L)}(\lambda,n), \tag{47}$$

$$\theta^{(K)}(\lambda,1) = \lambda^{(K)}. \tag{48}$$

Therefore (35) and (36) are correct for N+1. We have point out that they are correct for N=0,1, so they are correct for arbitrary N.

Substituting (35) and (36) into (16), we have

$$a_f^{(N)}(t) = \sum_\lambda \left[ b_{fi}^{(N)}(\lambda) + \sum_{n=1}^{N} \frac{1}{n!}\left(\frac{-it}{\hbar}\right)^n \sum_{M=n}^{N} b_{fi}^{(N-M)}(\lambda)\theta^{(M)}(\lambda,n) \right] \times \exp\left[\frac{i(E_f - \lambda^{(0)})t}{\hbar}\right]$$

(49)

### 1.3 The form of the Exact Solution of the Schrödinger Equation

According to the time-dependent perturbed approximation, the exact solution of the Schrödinger equation is

$$a_f(t) = \sum_{N=0}^{\infty} a_f^{(N)}(t)$$

(50)

We rewrite (49) as

$$a_f^{(N)}(t) = \sum_\lambda \left[ b_{fi}^{(N)}(\lambda) + \sum_{M=1}^{N} b_{fi}^{(N-M)}(\lambda)\sum_{n=1}^{M} \frac{1}{n!}\left(\frac{-it}{\hbar}\right)^n q^{(M)}(\lambda,n) \right] \times \exp\left[\frac{i(E_f - \lambda^{(0)})t}{\hbar}\right]$$

(51)

therefore

$$a_f(t) = \sum_\lambda \left[ \sum_{N=0}^{\infty} b_{fi}^{(N)}(\lambda) + \sum_{N=1}^{\infty} \sum_{M=1}^{N} b_{fi}^{(N-M)}(\lambda)\sum_{n=1}^{M} \frac{1}{n!}\left(\frac{-it}{\hbar}\right)^n \theta^{(M)}(\lambda,n) \right]$$
$$\times \exp\left[\frac{i(E_f - \lambda^{(0)})t}{\hbar}\right].$$

(52)

Suppose that

$$A = \sum_{K=0}^{\infty} A^{(K)}, \qquad B = \sum_{M=1}^{\infty} B^{(M)};$$

(53)

then we have

$$A \cdot B = \left[\sum_{K=0}^{\infty} A^{(K)}\right] \cdot \left[\sum_{M=1}^{\infty} B^{(M)}\right]$$
$$= A^{(0)}B^{(1)} + (A^{(1)}B^{(1)} + A^{(0)}B^{(2)}) + (A^{(2)}B^{(1)} + A^{(1)}B^{(2)} + A^{(0)}B^{(3)}) + \cdots$$
$$= \sum_{N=1}^{\infty} \sum_{M=1}^{N} A^{(N-M)}B^{(M)}.$$

(54)

Using this relation, (52) can be rewritten as

$$a_f(t) = \sum_\lambda \left\{ \sum_{N=0}^{\infty} b_{fi}^{(N)}(\lambda) + \sum_{K=0}^{\infty} b_{fi}^{(K)}(\lambda)\sum_{M=1}^{\infty} \left[ \sum_{n=1}^{M} \frac{1}{n!}\left(\frac{-it}{\hbar}\right)^n \theta^{(M)}(\lambda,n) \right] \right\}$$
$$\times \exp\left[\frac{i(E_f - \lambda^{(0)})t}{\hbar}\right]$$
$$= \sum_\lambda \left[ \sum_{N=0}^{\infty} b_{fi}^{(N)}(\lambda) \right]\left\{ 1 + \sum_{n=1}^{\infty} \frac{1}{n!}\left(\frac{-it}{\hbar}\right)^n \sum_{M=n}^{\infty} \theta^{(M)}(\lambda,n) \right\} \exp\left[\frac{i(E_f - \lambda^{(0)})t}{\hbar}\right]$$

Now we prove that

$$\left( \sum_{M=1}^{\infty} \lambda^{(M)} \right)^n = \sum_{M=n}^{\infty} \theta^{(M)}(\lambda, n).$$

(56)

From (48) we have

$$\sum_{M=1}^{\infty} \lambda^{(M)} = \sum_{M=1}^{\infty} \theta^{(M)}(\lambda, 1);$$

(57)

then (56) is true for n=1. Suppose (56) is correct for $n$. Therefore we have

$$\left( \sum_{M=1}^{\infty} \lambda^{(M)} \right)^{n+1} = \left( \sum_{L=1}^{\infty} \lambda^{(L)} \right) \cdot \sum_{M=n}^{\infty} \theta^{(M)}(\lambda, n)$$

$$= \sum_{K=n+1}^{\infty} \sum_{M=n}^{\infty} \theta^{(M)}(\lambda, n) \lambda^{(K-M)}$$

$$= \sum_{K=n+1}^{\infty} \theta^{(K)}(\lambda, n+1)$$

(58)

(56) is correct for $n+1$. So it is correct for arbitrary $n$. From (37) we have

$$\sum_{M=0}^{\infty} b_{fi}^{(M)}(\lambda) = \sum_{M=0}^{\infty} \sum_{L=0}^{M} c_f^{(L)}(\lambda) c_i^{(M-L)}(\lambda)^*$$

$$= \left( \sum_{L=0}^{\infty} c_f^{(L)}(\lambda) \right) \cdot \left( \sum_{M=0}^{\infty} c_i^{(M)}(\lambda)^* \right) = c_f(\lambda) c_i(\lambda)^*$$

(59)

Substituting (56) and (59) into (55), we finally obtain

$$a_f(t) = \sum_{I} c_f(I) c_i(I)^* \left\{ 1 + \sum_{n=1}^{\infty} \frac{1}{n!} \left[ \frac{-it}{\hbar} \left( \sum_{L=1}^{\infty} I^{(L)} \right) \right]^n \right\} \cdot \exp \left[ \frac{i(E_f - I^{(0)})t}{\hbar} \right]$$

$$= \sum_{I} c_f(I) c_i(I)^* \exp \left[ \frac{i(E_f - I)t}{\hbar} \right].$$

(60)

In this step it is necessary to verify that (60) is the solution of the Schrödinger equation (1) in the $g_1(t)$ case. In fact, according (23) we have

$$i\hbar \frac{da_f(t)}{dt} = \sum_{I} (I - E_f) c_f(I) c_i(I)^* \times \exp \left[ \frac{i(E_f - I)t}{\hbar} \right].$$

$$= \sum_{k} {}^{(S)}H'_{fk} \exp \left[ \frac{i(E_f - E_k)t}{\hbar} \right] a_k(t)$$

(61)

And satisfy the specific initial condition

$$a_f(t=0) = \delta_{fi}.$$

(62)

The result is affirmative, it is necessary to point out that (60) may be obtained in a direct way.

## 1.4 Transition Probability per Unit Time Deduced from the Exact Solution

According to quantum mechanics, the transition probability per unit time is defined as

$$w_{i \to f} = \lim_{t \to \infty} \sum_{E_l = E_f - \Delta E / 2}^{E_f + \Delta E / 2} \Delta E_l \, \boldsymbol{r}(E_l) a_l(t)^* a_l(t) / t$$

$$= \lim_{t \to \infty} \sum_{E_l = E_f - \Delta E / 2}^{E_f + \Delta E / 2} \Delta E_l \, \boldsymbol{r}(E_l) \sum_{\boldsymbol{n}} \sum_{\boldsymbol{l}} c_l(\boldsymbol{l})^* c_i(\boldsymbol{l}) c_l(\boldsymbol{n}) c_i(\boldsymbol{n})^* t^{-1} \qquad (63)$$

$$\times \exp\left[\frac{i(\boldsymbol{l} - \boldsymbol{n})t}{\hbar}\right] \quad .$$

Note that

$$\sum_{\boldsymbol{l}} c_l(\boldsymbol{l})^* c_l(\boldsymbol{l}) = 1. \qquad (64)$$

This means that $c_l(\boldsymbol{l})$ is the $\boldsymbol{l}$-component of one unit vector, and $c_i(\boldsymbol{l}) \exp(i\boldsymbol{l}t / \hbar)$ is the $\boldsymbol{l}$-component of another unit vector. The expression

$$\sum_{\lambda} c_l(\lambda)^* c_i(\lambda) \exp(i\lambda t / \hbar)$$

is the product of two unit vectors, so

$$\left| \sum_{\lambda} c_l(\lambda)^* c_i(\lambda) \exp(i\lambda t / \hbar) \right| \le 1. \qquad (65)$$

Finally we obtain

$$w_{i \to f} \le \lim_{t \to \infty} \sum_{E_l = E_f - \Delta E / 2}^{E_f + \Delta E / 2} \frac{\rho(E_l) \Delta E_l}{t} = 0. \qquad (66)$$

This is the result deduced from the exact solution of the Schrödinger equation (1) in the $g_1(t)$ case. This result cannot be used to describe the transition process.

## 2.5 Valid Region of the First Order Time-Dependent Approximation

The transition probability per unit time deduced from $a_f(t)$ cannot be used to describe the transition process, but the one deduced from $a_f^{(1)}(t)$ can be used to describe the transition process. Why?

From (50) to (60), we can see that the substance of the time-dependent perturbed approximation is to expand $\lambda$, $c_f(\lambda)$ presented in (60) according to the stationary-state perturbed approximation. Therefore, the convergence intervals of the time-dependent perturbed approach and the stationary-state perturbed approach are identical. We know that the convergence interval of the stationary-state perturbed approach is

$$\left| {}^{(S)}H'_{fi} \right| \le \left| E_i - E_f \right| \tag{67}$$

So it must be the convergence interval of the time-dependent perturbed approach. Let us examine the first-order approximation of the time-dependent perturbed approximation

$$a_f^{(1)}(t)\Big|_{f \ne i} = \sum_l \left[ c_f^{(1)}(\lambda^{(0)} = E_l)c_i^{(0)}(\lambda^{(0)} = E_l)^* + c_f^{(0)}(\lambda^{(0)} = E_l)c_i^{(1)}(\lambda^{(0)} = E_l)^* \right]$$

$$\times \exp\left[ \frac{i(E_f - \lambda^{(0)})t}{\hbar} \right]$$

$$= c_f^{(1)}(\lambda^{(0)} = E_i)\exp\left[ \frac{i(E_f - E_i)t}{\hbar} \right] + c_i^{(1)}(\lambda^{(0)} = E_f)^*$$

$$= \frac{{}^{(S)}H'_{fi}}{E_i - E_f}\left\{ \exp\left[ \frac{i(E_f - E_i)t}{\hbar} \right] - 1 \right\} \tag{68}$$

$c_f(\lambda)$ is the $l$-component of a unit vector. Then

$$\left| c_f^{(1)}(\lambda^{(0)} = E_i) \right| = \left| \frac{{}^{(S)}H'_{fi}}{E_i - E_f} \right| \le 1. \tag{69}$$

and (68) can only be used in the region $\left| {}^{(S)}H'_{fi} \right| \le \left| E_i - E_f \right|$.

From quantum mechanics we have

$$w_{i \to f}^{(1)} = \sum 4\left| {}^{(S)}H'_{fi} \right|^2 \frac{\sin^2\left[ (E_f - E_i)t/(2\hbar) \right]}{(E_f - E_i)^2} \rho(E_f)\Delta E_f \tag{70}$$

$$\to \int 4\left| {}^{(S)}H'_{fi} \right|^2 \frac{\sin^2\left[ (E_f - E_i)t/(2\hbar) \right]}{(E_f - E_i)^2} \rho(E_f)dE_f$$

$$= \left( \frac{2\pi t}{\hbar} \right){}^{(S)}H'_{fi}\Big|^2 \rho(E_f). \tag{71}$$

First, the premise of this section is that we discuss the discrete case. Quantum mechanics changes to discuss the continuous case and substitutes (70) with (71). Is this reasonable? We shall discuss this question in the next section. Second, the transition probability deduced from the perturbed approximation is correct only in the region $\left| {}^{(S)}H'_{fi} \right| < \left| E_f - E_i \right|$; the fault is in $\left| E_f - E_i \right| < \left| {}^{(S)}H'_{fi} \right|$.

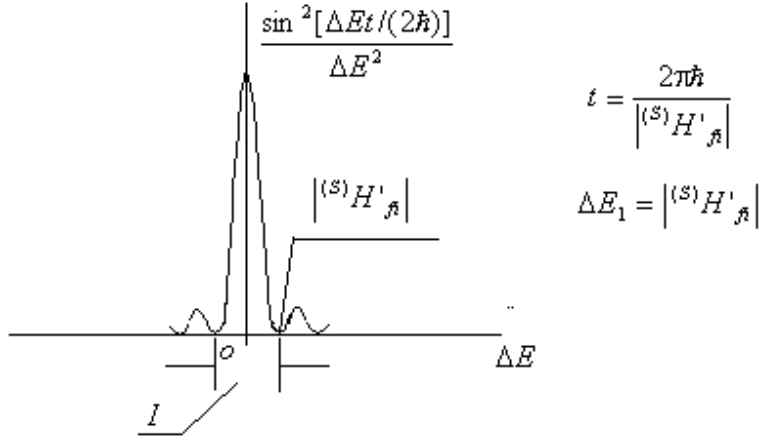

Fig. 1  The relation between  $\sin^2[\Delta E t/(2\hbar)]/(\Delta E)^2$  and  $\Delta E = E_f - E_i$.

In Fig. 1,  $I$  is the divergent region of the perturbed approximation series. The first zero of this curve near the main peak is at  $\dfrac{|\Delta E_1|t}{2\hbar} = \pi$. If  $t = \dfrac{2\pi\hbar}{\left|{}^{(S)}H'_{fi}\right|}$, then we have  $\left|\Delta E_1\right| = \left|{}^{(S)}H'_{fi}\right|$. And the complete main pear is in the divergent region of the perturbed approximation series. The problem is that it is under these two conditions,  $t \geq \dfrac{2\pi\hbar}{\left|{}^{(S)}H'_{fi}\right|}$, and  $\left|\Delta E\right| \leq \left|{}^{(S)}H'_{fi}\right|$, that quantum mechanics obtains Fermi's golden rule

$$w_{i\to f} = \frac{2\pi}{\hbar}\left|{}^{(S)}H'_{fi}\right|^2 \rho(E_f) \tag{72}$$

from (71). Of course, if this formula is correct, then others can find a suitable  ${}^{(S)}H'_{fi}$  to be identical to the experimental value  $w_{i\to f}$, but mathematically Fermi's rule is not a reasonable conclusion from the Schrödinger equation.

We summarize the discussion in the discrete case:

(1) The substance of the time-dependent perturbed approximation is to expand  $\lambda$,  $c_l(\lambda)$  presented in (60) according to the stationary-state perturbed approach.

(2) The convergence interval of the stationary-state perturbed approach is  $\left|{}^{(S)}H'_{fi}\right| \leq \left|E_f - E_i\right|$, which is also the convergence interval of the time-dependent perturbed approach.

(3) The sum of the first finite-order perturbed approximate solutions may be used to substitute to the exact solution only in the convergence interval. The fundamental error of the time-dependent perturbed approach is using the sum of the first finite-order approximate solutions to calculate the transition probability in the interval  $\left|{}^{(S)}H'_{fi}\right| > \left|E_f - E_i\right|$.

Mathematically Fermi's golden rule is not a reasonable conclusion from the Schrödinger equation in the discrete case.

(4) People use the sum of first finite order perturbed approach to substitute the exact solution. It only caused by that they cannot find out the approximate solution $a_f^{(N)}(t)$ for arbitrary $N$.

When we find out $a_f^{(N)}(t)$ and $a_f(t) = \sum_{N=0}^{\infty} a_f^{(N)}(t)$, then we have not any reason to use the sum of first finite order perturbed approach to substitute the exact solution. However the transition probability per unit time deduced from the exact solution the Schrödinger equation is zero in the discrete case; it cannot be used to describe the transition process.

## 2. DISCUSSION OF THE CASE IN WHICH THE TIME FACTOR IS $g_1(t)$ AND THE EIGENVALUES OF $\hat{H}_0$ ARE CONTINUOUS

In the continuous case there are two ways to treat the problem. One way is to assume that the system is contained in a large cubical box of dimensions L that has periodic boundary conditions at its walls. Then the eigenvalues become discrete, and the above-mentioned result can be utilized under this condition.

Another way is proceed with $E_k$ as a continuous variable and to change the summation into the integration, and $\delta_{fi}$ presented in the discrete case must be changed into $\delta(E_f - E_i)$. In the following we shall emphasize the distinction of both cases.

### 3.1 Exact Solution

In the continuous case the Schrödinger equation has the form

$$i\hbar \frac{da_f(t)}{dt} = \int dE_k {}^{(S)}H'_{fk} \exp\left[\frac{i(E_f - E_k)t}{\hbar}\right] a_k(t) \tag{73}$$

The initial condition can be expressed in the form

$$a_f(t=0) = A\boldsymbol{\delta}(E_f - E_i). \tag{74}$$

Detailed discussion shows that the form of $\delta_{fi}$ on the right-hand side of Eq. (74) is no longer applicable in the continuous case. Under the circumstances, if the initial condition takes the form $a_f(t=0) = \delta_{fi}$, then at $t=0$ the probability of the system in the region $(E_f - \Delta E/2, \ E_f + \Delta E/2)$ is

$$\int_{E_f - \Delta E/2}^{E_f + \Delta E/2} dE_I a_I(t=0)^* a_I(t=0) = 0,$$

where $E_f$ is arbitrary. Obviously it is incorrect. If the initial condition takes the form of (74), then at $t = 0$ the probability of the system in the same region is

$$\int_{E_i - \Delta E/2}^{E_i + \Delta E/2} a_l(t=0)^* a_l(t=0) dE_l$$

$$= |A|^2 \int_{E_i - \Delta E/2}^{E_i + \Delta E/2} \delta(E_l - E_i)\delta(E_l - E_i) dE_l = |A|^2 \delta(0). \tag{75}$$

Since only the relative transition probability per unit time is calculated, the existence of the term $\delta(0)$ would not have any impact on our discussion. Therefore, the initial condition must take the form of (74).

We can prove that the exact solution of the Schrödinger equation (43) and the initial condition is

$$a_f(t) = A \int d\lambda c_f(\lambda) c_i(\lambda)^* \exp\left[\frac{i(E_f - \lambda)t}{\hbar}\right]. \tag{76}$$

where $\lambda$, $c_f(t)$ are the eigenvalue and eigenfunction of the stationary-state equation, respectively,

$$\int dE_k [E_f \delta(E_f - E_k) + {}^{(S)}H'_{fk}] c_k(\lambda) = \lambda c_f(\lambda) \tag{77}$$

At the same time, the eigenfunctions satisfy the orthonormality conditions

$$\int d\lambda c_f(\lambda) c_i(\lambda)^* = \delta(E_f - E_i) \tag{78}$$

$$\int dE_j c_j(\boldsymbol{\lambda}) c_j(\boldsymbol{\eta})^* = \boldsymbol{\delta}(\boldsymbol{\lambda} - \boldsymbol{\eta}). \tag{79}$$

## 2.2 Equivalence of the Time-Dependent Perturbed Approximation and the Stationary-State Perturbed Approximation

The stationary-state perturbed approximation (77) is

$$\lambda = \sum_{N=0}^{\infty} \lambda^{(N)}, \qquad\qquad c_j(\lambda) = \sum_{N=0}^{\infty} c_j^{(N)}(\lambda), \tag{80}$$

where $\lambda^{(N)}$, $c_j^{(N)}(\lambda)$ satisfy a set of equations

$$(\lambda^{(0)} - E_f)c_f^{(0)} = 0,$$

$$(81)$$

$$(\lambda^{(0)} - E_f)c_f^{(N)}(\lambda) + \sum_{L=1}^{N} \lambda^{(L)} c_f^{(N-L)}(\lambda) = \int dE_k^{(S)} H'_{fk} c_k^{(N-1)},$$

and satisfy the orthonormality condition

$$\int d\lambda c_f^{(0)}(\lambda)c_i^{(0)}(\lambda)^* = \delta(E_f - E_i) \tag{82}$$

$$\int d\lambda \sum_{L=0}^{N} c_f^{(L)}(\lambda)c_i^{(N-L)}(\lambda)^* = 0. \tag{83}$$

Inserting (80) into (76) we obtain the expansion of exact solution in terms of the stationary-state perturbed approach

$$a_f(t) = \sum_{N=0}^{\infty} a_f^{(N)}(t), \tag{84}$$

where

$$a_f^{(N)}(t) = A \int d\lambda \left[ b_{fi}^{(N)}(\lambda) + \sum_{n=1}^{N} \frac{1}{n!} \left( \frac{-it}{\hbar} \right)^n \sum_{L=n}^{N} b_{fi}^{(N-L)}(\lambda) \theta^{(L)}(\lambda, n) \right] \exp\left[ \frac{i(E_f - \lambda^{(0)})t}{\hbar} \right],$$

$$(85)$$

and $b_{fi}^{(N)}(\lambda)$, $\theta^{(L)}(\lambda, n)$ are defined by (37), (47) and (48), respectively.

In the same manner that is utilized in section 2.1, we can prove that $a_f^{(N)}(t)$ satisfy the time-dependent perturbed approximation equation

$$i\hbar \frac{da_f^{(0)}(t)}{dt} = 0,$$

$$i\hbar \frac{da_f^{(N)}(t)}{dt} = \int dE_k^{(S)} H'_{fk} \exp\left[ \frac{i(E_f - E_k)t}{\hbar} \right] a_k^{(N-1)}. \tag{86}$$

So the time-dependent perturbed approximation is equivalent to the stationary-state perturbed approximation in this case.

## 2.3 Error of Deduction of Transition Probability from the Perturbed Approach

The zeroth and first-order stationary-state perturbed approximations are

$$c_f^{(0)}(\lambda^{(0)} = E_i) = \delta(\lambda^{(0)} - E_f) = \delta(E_i - E_f), \tag{87}$$

$$c_f^{(1)}(\lambda^{(0)} = E_i) =^{(S)} H'_{fi} /(E_i - E_f). \tag{88}$$

Comparing $c_f^{(0)}(\lambda^{(0)} = E_i)$ and $c_f^{(1)}(\lambda^{(0)} = E_i)$, we can see that

$$\frac{\left|c_f^{(1)}(\lambda^{(0)}=E_i)\right|}{\left|c_f^{(0)}(\lambda^{(0)}=E_i)\right|_{\max}}=\frac{\left|^{(S)}H'_{fi}\right|}{\left|E_i-E_f\right|}\cdot\frac{\hbar\pi}{t}>1 \tag{89}$$

where $\left|\Delta E\right|t\to0$ in the neighborhood of the main peak of $\delta$ function. So we cannot use the sum of first finite order perturbed approach to substitute the exact solution in this case. However there are still two other mistakes.

First, even if we utilize the first-order approximation, the expression

$$\int\limits_{E_f-\Delta E/2}^{E_f+\Delta E/2}dE_l\rho(E_l)t^{-1}a_l^{(1)}(t)a_l^{(1)}(t)^*=\frac{2\pi}{\hbar}\left|^{(S)}H'_{fi}\right|^2\rho(E_f) \tag{90}$$

would not be the relative transition probability per unit time, because at $t=0$ the probability of the system in $(E_i-\Delta E/2,\ E_i+\Delta E/2)$ is $\left|A\right|^2\delta(0)=1$. Fermi's golden rule is not a reasonable inference. The relative transition probability peer unit time is

$$w_{i\to f}^{(1)}=\frac{2\pi\left|A\right|^2}{\hbar}\left|^{(S)}H'_{fi}\right|^2\rho(E_f)=\frac{2\pi}{\hbar\delta(0)}\left|^{(S)}H'_{fi}\right|^2\rho(E_f). \tag{91}$$

It still cannot be applied to describe the time-dependent process.

Second, if we only take the zeroth and the first-order approximations into account, which means that

$$c_f^{(N)}(\lambda)=0,\quad if\quad N\ge2, \tag{92}$$

then together with (87) and (88) the second-order orthonormality condition

$$\int d\lambda[c_f^{(0)}(\lambda)c_i^{(2)}(\lambda)^*+c_f^{(1)}(\lambda)c_i^{(1)}(\lambda)^*+c_f^{(2)}(\lambda)c_i^{(0)}(\lambda)^*]=0 \tag{93}$$

cannot be satisfied.

## 2.4 Deduction of the Transition Probability from the Exact Solution

The transition probability in $(0,\ t)$ is a relative probability. It is defined as

$$\begin{aligned}W(t)&=\frac{the\ probability\ at\ t\ in\ (E_f-\Delta E/2,\ E_f+\Delta E/2)}{the\ probability\ at\ t=o\ in\ all\ energy\ regious}\\&=\frac{\displaystyle\int\limits_{E_f-\Delta E/2}^{E_f+\Delta E/2}a_l^*(t)a_l(t)\rho(E_l)dE_l}{\displaystyle\int\limits_{all\ energy}a_l^*(0)a_l(0)dE_l}\end{aligned} \tag{94}$$

In this equation the denominator is

$$\int\limits_{all\ energy} a_l(t=0)^* a_l(t=0)dE_l$$

$$= |A|^2 \int\limits_{E_i - \Delta E/2}^{E_i + \Delta E/2} \delta(E_l - E_i)\delta(E_l - E_i)dE_l = |A|^2 \delta(0); \tag{95}$$

the numerator is

$$\int\limits_{E_f - \Delta E/2}^{E_f + \Delta E/2} \rho(E_l)dE_l a_l^*(t)a_l(t)$$

$$= |A|^2 \int\limits_{E_f - \Delta E/2}^{E_f + \Delta E/2} \rho(E_l)dE_l \int c_l(\lambda)c_i^*(\lambda)\exp\left[\frac{-i\lambda t}{\hbar}\right]d\lambda \int c_l^*(\nu)c_i(\nu)\exp\left[\frac{i\nu t}{\hbar}\right]d\nu \tag{96}$$

In the continuous case, $c_l(\boldsymbol{l})$ satisfies the orthonormality condition (78). Then $c_l(\boldsymbol{l})$ is the $\boldsymbol{l}$-component of the normality vector, and $c_i^*(\boldsymbol{l})\exp(-i\boldsymbol{l}t/\hbar)$ is the $\boldsymbol{l}$-component of another normality vector. The expression

$$\int d\lambda c_l(\lambda)c_i^*(\lambda)\exp\left(\frac{-i\lambda t}{\hbar}\right)$$

is the product of two normality vectors, so

$$\left|\int d\lambda c_l(\lambda)c_i^*(\lambda)\exp\left(\frac{-i\lambda t}{\hbar}\right)\right| \leq \delta(E_l - E_i) \tag{97}$$

The transition probability per unit time is

$$w_{i\to f} = \lim_{t\to\infty} \frac{W(t)}{t}$$

$$\leq \lim_{t\to\infty}\left(\frac{1}{t\boldsymbol{d}(0)}\right)\int\limits_{E_f - \Delta E/2}^{E_f + \Delta E/2} \boldsymbol{r}(E_l)dE_l\boldsymbol{d}(E_l - E_i)\boldsymbol{d}(E_l - E_i) = \lim_{t\to\infty}\frac{\boldsymbol{r}(E_f)}{t} = 0. \tag{98}$$

In the continuous case the transition probability per unit time deduced from the exact solution of the Schrödinger equation is still zero.

## 3. DISCUSSION OF THE CASE IN WHICH THE TIME FACTOR IS $g_2(t)$ AND THE EIGENVALUES OF $\hat{H}_0$ ARE DISCRETE

Quantum mechanics uses

$$\hat{H}'(t) = g_2(t)\hat{B} = \hat{B}\cos(\boldsymbol{w} + \boldsymbol{d}), \qquad 0 < t. \tag{99}$$

to describe the electron transition from one energy level to another due to the impact of the electromagnetic field. In general, only the discrete is discussed. The Schrödinger equation has the form

$$i\hbar \frac{da_f(t)}{dt} = \sum_k \left\{ F_{fk} \exp\left[\frac{i(E_f + \hbar\omega - E_k)t}{\hbar}\right] + F_{fk}^* \exp\left[\frac{i(E_f - \hbar\omega - E_k)t}{\hbar}\right] \right\} a_k(t) \quad (100)$$

where

$$F_{fk} = B_{fk} e^{ib} / 2 . \qquad (101)$$

Suppose

$$a_f(t) = \sum_{L=0}^{\infty} A_{fL}(t) \qquad (102)$$

where $A_{fL}(t)$ is the probability amplitude of the state in which there is an electron in $E_f$ and $L$ photons of energy $\hbar\omega$. If at $t = 0$ the system is in the state $(E_{i,} \ L_0 \hbar\omega)$, then

$$A_{fL}(t = 0) = \delta_{fi}\delta_{LL_0} . \qquad (103)$$

From (100) and (102) we get

$$A_{fL}(t) = \sum_{m} c_{fL}(m) c_{iL_0}(m)^* \exp\left[\frac{i(E_f + L\hbar\omega - m)t}{\hbar}\right], \qquad (104)$$

where $c_{fL}(\mu)$ is the solution of the stationary-state equation

$$(\mu - E_f - L\hbar\omega)c_{fL}(\mu) = \sum_k \left[ F_{fk}c_{k,L-1}(\mu) + F_{fk}^* c_{k,L+1}(\mu) \right] \qquad (105)$$

Suppose

$$K_{fL,kL'} = \begin{cases} F_{fk} & if \ L' = L-1 \\ F_{fk}^* & if \ L' = L+1, \\ 0 & if \ L' \neq L \pm 1 \end{cases} \qquad (106)$$

then (105) may be rewritten in the form

$$m c_{fL}(m) = \sum_k \sum_{L'} \left[ (E_f + L\hbar\omega)\delta_{fk}\delta_{LL'} + K_{fL,kL'} \right] c_{kL'}(m) . \qquad (107)$$

At the same time, $c_{fL}(m)$ meets the orthonormality conditions of the eigenfunctions:

$$\sum_{v} c_{fL}(v)c_{iL'}^*(v) = \delta_{fi}\delta_{LL'} \qquad (108)$$

$$\sum_{j} \sum_{L} c_{jL}(\lambda)c_{jL}^*(v) = \delta_{\lambda v} \qquad (109)$$

Obviously, the exact solutions solved under the cases $g_1(t)$ and $g_2(t)$ have the identical mathematical structure. The results of Sec.2 are still applicable in this case.

## 4. CONCLUSION

(1) In the cases in which the time factor are $g_1(t)$ or $g_2(t)$, the eigenvalues of $\hat{H}_0$ are discrete or continuous; the processes to deduce the transition probability per unit time from time-dependent perturbed approximations contains a basic mathematical mistake. Fermi's golden rule is not the mathematically deductive inference from the Schrödinger equation.

(2) The transition probability per unit time deduced from the exact solution of the Schrödinger equation cannot be used to describe time-dependent processes.

Welcome to www.newpo,com
Email: junhaoz@pub.shantou.gd.cn